\def\cm{{\rm\thinspace cm}}

\def\erg{{\rm\thinspace erg}}

\def\keV{{\rm\thinspace keV}}
\def\km{{\rm\thinspace km}}

\def\Mpc{{\rm\thinspace Mpc}}
\def\Msun{\hbox{$\thinspace M_{\odot}$}}

\def\s{{\rm\thinspace s}}
\def\yr{{\rm\thinspace yr}}
\def\sr{{\rm\thinspace sr}}

\def\ergpcmsqps{\hbox{$\erg\cm^{-2}\s^{-1}\,$}}

\def\ergps{\hbox{$\erg\s^{-1}\,$}}

\def\kmps{\hbox{$\km\s^{-1}\,$}}

\def\psqcm{\hbox{$\cm^{-2}\,$}}

\def\kmpspMpc{\hbox{$\kmps\Mpc^{-1}$}}

\def\lax{{$\mathrel{\hbox{\rlap{\hbox{\lower4pt\hbox{$\sim$}}}\hbox{$<$}}}$}}
\def\gax{{$\mathrel{\hbox{\rlap{\hbox{\lower4pt\hbox{$\sim$}}}\hbox{$>$}}}$}}

\def\aa{{A\&A}}

\def\aj{{AJ}}

\def\apj{{ApJ}}

\def\mnras{{MNRAS}}
\def\nat{{Nature}}

\documentclass[cup5b]{caps}
\usepackage{graphicx}
\usepackage{amssymb}
\usepackage{ociwsymp1}
\HeadText{A. C. Fabian}

\begin{document}

\pagenumbering{arabic}

\author[]{ANDREW C. FABIAN\\Institute of Astronomy, Cambridge, UK}

\chapter{Obscured Active Galactic Nuclei \\ and Obscured Accretion}

\begin{abstract}
Most of the local active galactic nucleus (AGN) population is obscured and 
much of the X-ray background originates in obscured AGNs. The contribution of 
obscured accretion to the growth of massive black holes is discussed here. 
The recent identification of significant samples of the X-ray sources
that dominate the X-ray background intensity has shown a redshift
peak at 0.7--0.8, rather than the redshift of 2 found for bright
optical quasars. Obscured accretion has a faster evolution than
unobscured accretion. The lower redshift and luminosity of most
obscured AGNs mean that although they dominate the absorption-corrected
intensity of the X-ray background by a factor of about 3 over
unobscured objects, they make only an equal contribution to the local
mass density in black holes. Obscured and unobscured AGNs together
contribute about $4\times 10^5\Msun\Mpc^{-3}$. Type 2 quasars and
Compton-thick objects may give another $10^5\Msun\Mpc^{-3},$ but no
more unless direct determinations from the $M_{\bullet}-\sigma$ relation
seriously underestimate the local black hole mass density, or unless
most massive black holes are rapidly spinning (so having a higher
radiative efficiency than the 10\% assumed above). Obscured accretion probably 
dominates the growth of black holes with masses below a few times $10^8\Msun$, 
whereas optically bright quasars dominate at higher masses. The luminosity 
absorbed by the dusty gas in obscured AGNs is reradiated in the mid-infrared 
and far-infrared bands. The contribution of AGNs drops from about 20\% of the 
mid-infrared background to just a few percent of the far-infrared background.

\end{abstract}

\section{Introduction}
The X-ray background (XRB) is dominated by the emission from active galactic 
nuclei (AGNs). This enables a census to be made of the radiative growth of 
massive black holes. The infrared and sub-mm backgrounds (hereafter IRB) are 
dominated by emission from star formation.  Together the backgrounds provide 
measures of the evolution of black holes and galaxies.

The situation is complicated, however, by the fact that the bulk of the XRB 
is due to highly obscured AGNs. This was first predicted by Setti \& Woltjer 
(1989), elaborated on by Madau, Ghisellini, \& Fabian (1994), Comastri et al. 
(1995), Gilli, Risaliti, \& Salvati (1999), and others, and demonstrated by 
direct resolving of the XRB with {\it Chandra}\ by Mushotzky et al. (2000), 
Brandt et al. (2001), Giacconi et al. (2001), and Rosati et al. (2002), and  
with {\it XMM-Newton}\ by Hasinger et al. (2001). Simple 
pre-{\it Chandra}/{\it XMM-Newton}\ estimates (Fabian \& Iwasawa 1999), based 
on a comparison of the intensity of the 2--10~keV XRB, which is dominated by
obscured AGNs with that of the soft XRB below 1~keV, which is dominated by
unobscured quasars, indicated that most accretion may be obscured (i.e., 
occurring behind a line-of-sight column density exceeding 
$N_{\rm H}=10^{22}\psqcm$). This assumed that the redshift evolution of 
obscured and unobscured objects is the same. Recent {\it Chandra}\ and 
{\it XMM-Newton}\  data, however, show that this is not the case.

That obscured AGNs are common and need to be included in estimates of 
accretion power is obvious from the fact that the three nearest AGNs with 
intrinsic X-ray luminosities above $10^{40}\ergps$ (NGC 4945, the Circinus 
galaxy, and Centaurus A) are all highly obscured with 
$N_{\rm H}>10^{23}\psqcm$ (Matt et al. 2000). Two (NGC 4945 and Circinus) are 
even Compton-thick with $N_{\rm H}>1.5\times 10^{24}\psqcm.$ This situation 
has only been slowly appreciated, perhaps due to NGC 4945 appearing as 
a starburst galaxy at all non-X-ray wavelengths and to the Circinus galaxy 
lying close to the Galactic plane. 

Nevertheless, it has long been known that the number density of
Seyfert 2 galaxies, where the active nucleus is obscured, exceeds
that of Seyfert 1 galaxies, although selection effects complicate
making a comparison at a fixed bolometric AGN luminosity. Geometrical
unification has often been assumed, with Seyfert 2s being Seyfert 1s
viewed through a surrounding torus for which the opening angle is
about 60$^{\circ}$. More recent X-ray studies, particularly with 
{\it BeppoSAX}, have been showing that this picture probably applies to only 
a subset of AGNs.

Here, the evidence for distant obscured AGNs is reviewed and their
contribution to the XRB examined. The total energy density due to
accretion is then deduced from the spectrum of the XRB, and via
So\l tan's (1982) method converted into a local black hole mass density
due to radiative growth. Comparison with the locally determined black
hole mass density from quiescent galaxies shows that there could be a
problem in terms of excessive growth, unless (1) the radiative
efficiency of most accretion is higher than that for a standard accretion
disk around a Schwarzschild (non-spinning) black hole (e.g., Elvis,
Risaliti, \& Zamorani 2002), (2) the bolometric correction is lower, or 
(3) the redshift distribution peaks at $z<2$.

The problem is illustrated well by the recent estimate for the growth
of bright optical quasars by Yu \& Tremaine (2002), which allows for
little obscured accretion, particularly in massive objects. It is
shown here that if their estimate for the local black hole mass
density from direct measurements of nearby quiescent galaxies can be
revised upward by a factor of 1.5--2, to be in agreement with that of
Ferrarese (2002), and factors due to (2) and (3) above are also revised
in accord with recent XRB studies, then agreement can be found for a
radiative efficiency of 0.1. About equal amounts of the local black
hole mass density are then due to obscured and to unobscured
accretion.

The X-ray and UV energy absorbed in the obscuring gas is reradiated in the 
far-infrared and sub-mm bands. A few percent of the energy density in these 
backgrounds is due to accretion, but most is due to star formation. 

\section{Obscured AGNs}

Most Seyfert 2 galaxies contain obscured AGNs, with 2--10~keV X-ray
luminosities typically up to about $10^{44}\ergps$. In studies of local,
optically selected Seyfert 2s with {\it BeppoSAX}, Maiolino et al.
(1998) have shown that about one-half are Compton thick. In general
this half are the classical, optical Seyfert 2 galaxies and the
other, Compton-thin, half corresponds to the optical intermediate
classes of Seyfert 1.8 and 1.9 (Risaliti, Maiolino, \& Salvati 1999).  

Distant obscured AGNs (redshift $z>0.3$) are now being found in large numbers 
by X-ray observations with {\it Chandra}\ and {\it XMM-Newton}\ (Mushotzky
et al. 2000; Alexander et al. 2001; Barger et al. 2001; Brandt et al. 2001;
Crawford et al. 2001; Giacconi et al. 2001; Hasinger et al. 2001;
Rosati et al. 2002). Most of the serendipitous sources found in an
X-ray image above 1~keV made with these telescopes are obscured.
Source variability in many cases makes an AGN identification
unambiguous. The determination of column densities requires that the
source is identified and its redshift known. Where significant samples
are available (e.g., Alexander et al. 2001; Barger et al. 2002; Mainieri
et al. 2002), more than two-thirds are Compton thin with
$10^{21}<N_{\rm H}<10^{23}\psqcm$; most of the remainder are
unobscured. The absorption-corrected, 2--10~keV luminosity of the
obscured objects is typically in the range of $10^{42}-10^{44}\ergps$.
Only a handful of obscured AGNs have yet been found with 2--10~keV
luminosities exceeding the level of $\sim3\times 10^{44}\ergps$,
corresponding to a quasar [Crawford et al. 2002; Norman et al. 2002; 
Stern et al. 2002; Wilman et al. 2003; Mainieri et al. 2002; the last
authors define a quasar as $L(0.5-10\keV)>10^{44}\ergps$].

Note that there is not complete agreement between optical and X-ray
classification of some of the Compton-thin AGNs. Some show X-ray absorption 
but little optical extinction (Maiolino et al. 2001), and vice versa, and some 
X-ray obscured objects show no detectable narrow-line region at optical or 
infrared wavelengths (e.g., Comastri et al. 2002; Gandhi, Crawford, \& Fabian 
2002). When there is a large covering fraction of the nucleus by dusty gas, 
there need be little or no optical/UV narrow-line region. The terms Type 2 and 
Type 1 when applied to X-ray sources are commonly referring to whether there 
is absorption or not, irrespective of the optical spectrum.

\begin{figure}
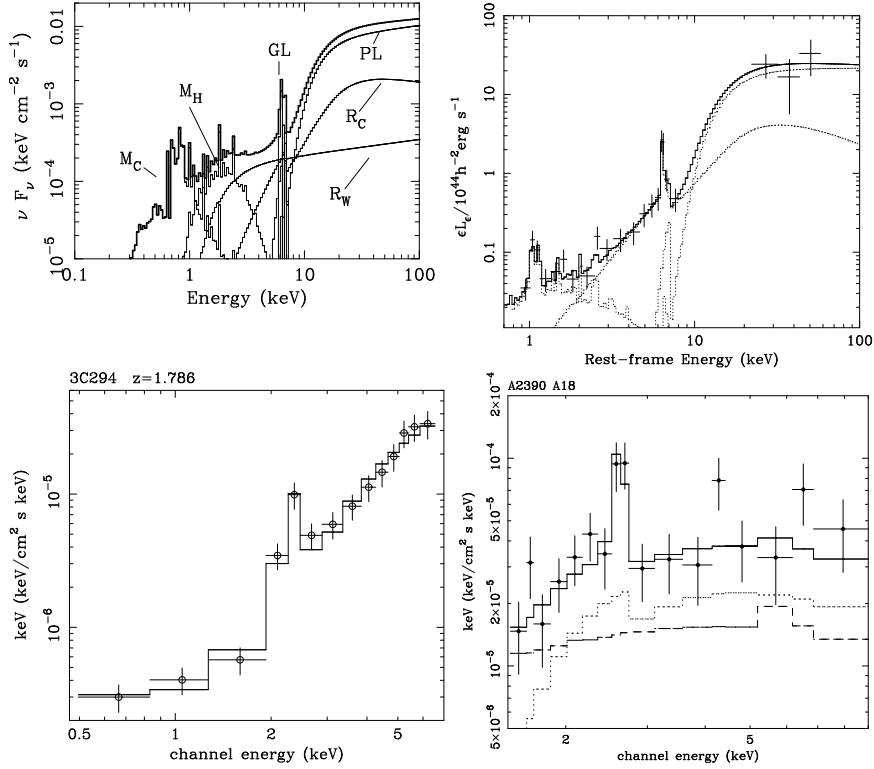

{\includegraphics[width=.35\columnwidth,angle=-90]{n6240.ps} \hfil
\includegraphics[width=.42\columnwidth,angle=-90]{ir09xed.ps}}
{\includegraphics[width=.45\columnwidth,angle=-90]{3c294_eeuf.eps} \hfil
\includegraphics[width=.45\columnwidth,angle=-90]{a18bwa.ps}}
\caption{Examples of the $\nu F_{\nu}$ X-ray spectra of obscured AGNs.
{\it Top left:}\ Model spectrum that fits the {\it BeppoSAX}\ data on
NGC 6240 (Vignati et al. 1999). The heavily absorbed power law (PL)
and Gaussian line (GL) emission of iron are shown, with hot and cold
reflection and emission components. {\it Top right:}\ IRAS 09104+4109
has a 2--10~keV luminosity of $\sim 10^{46}\ergps$ behind a column
density of $3\times 10^{24}\psqcm$ (Franceschini et al. 2000; Iwasawa,
Fabian, \& Ettori 2001). {\it Lower left:}\ is 3C 294, a powerful
radio galaxy at $z=1.786$ with an X-ray luminosity of $\sim
10^{45}\ergps$ and a column density of $8\times 10^{23}\psqcm$. {\it
Lower right:}\ is an {\it XMM-Newton}\ spectrum (Gandhi 2002) of
serendipitous source A18 at $z=1.467$ in the field of the rich galaxy
cluster A2390 that has a 2--10~keV luminosity of $\sim 10^{45}\ergps$
and $N_{\rm H}\approx 2\times 10^{23}\psqcm$. The level of
contaminating cluster emission to the spectrum is indicated by the
dashed line.}
\end{figure}

Another class of obscured AGNs are the powerful radio galaxies. These
have large column densities (probably in a torus perpendicular to the
radio axis) of $\sim 10^{23}\psqcm$ or more (e.g., Cygnus A, Ueno et al
1994; 3C 294 at $z=1.786$, Fabian et al. 2003; B2 0902 at $z=3.2$,
Fabian, Crawford, \& Iwasawa 2002; see Fig. 1.1). These are sufficiently rare 
that their contribution to the XRB intensity is negligible. 

Source counts from deep X-ray surveys show that most of the XRB is now
resolved (Fig. 1.2), with a major uncertainty being the actual intensity of the
XRB measured by wide-beam instruments. The counts flatten below a flux
of about $10^{-14}\ergpcmsqps$ in the 2--10~keV band, above which more
than 60\% of the XRB intensity originates. At much lower fluxes
more and more starburst galaxies are detected (Alexander et al. 2002a;
Hornschemeier et al.  2002). They make a negligible contribution to
the total XRB intensity (Brandt et al. 2002).

\begin{figure}
\centerline{
\includegraphics[width=\columnwidth,angle=-90]{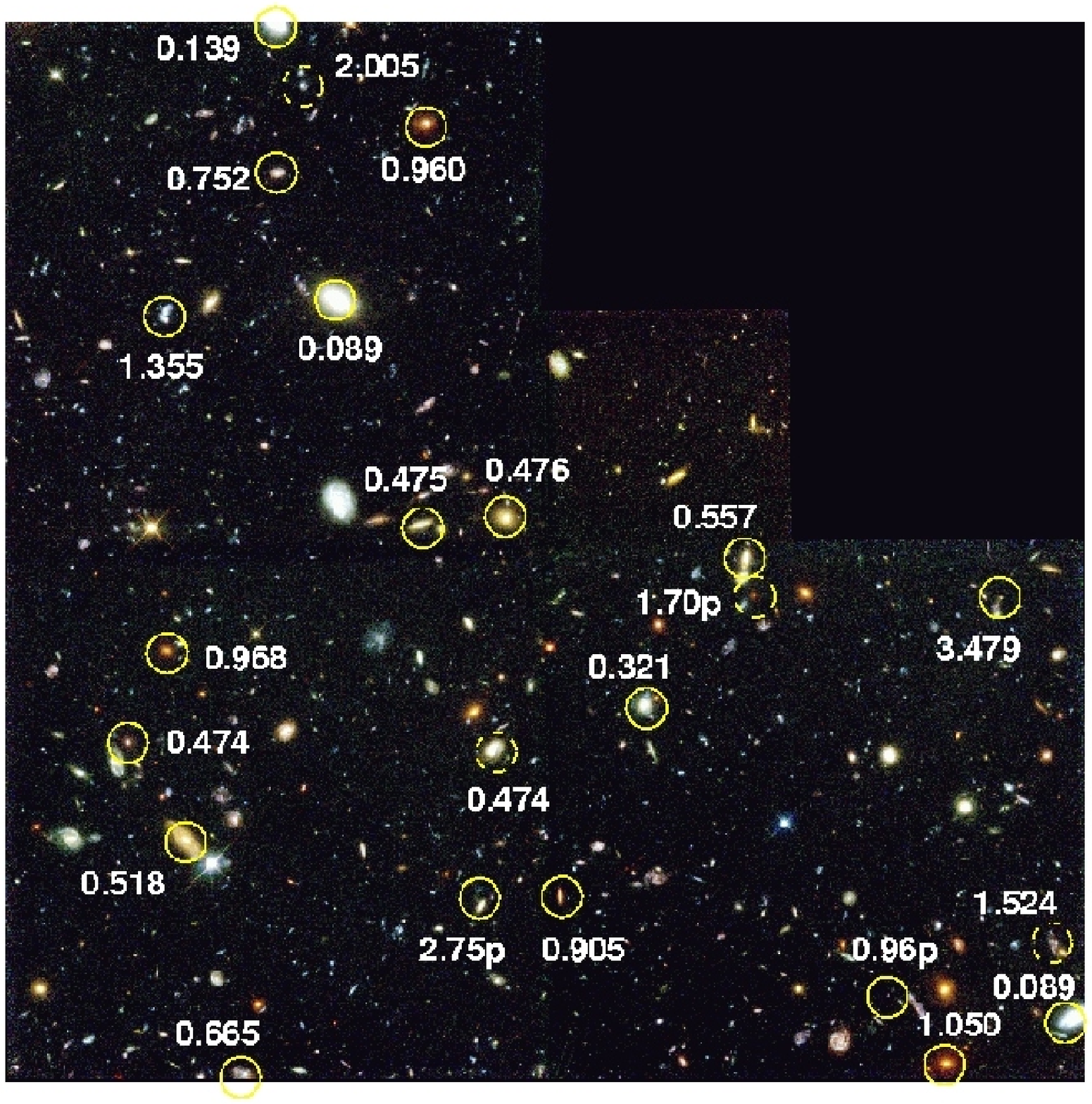}}
\caption{{\it Chandra}\ sources from a 2~Ms exposure of the HDF-N superposed
on a {\it Hubble Space Telescope}\ optical image (kindly provided by W. N. 
Brandt; see, e.g., Brandt et al. 2002). Source redshifts are indicated 
(``p'' means photometric). }
\end{figure}

\section{X-ray Constraints on the Radiative Growth of Massive Black Holes}

The basic method for deducing the local density in black holes due to
growth by accretion that emitted measurable radiation is that
originally due to So\l tan (1982). From $$E=\eta M c^2,$$ where $\eta$
is the efficiency with which mass is turned into radiation
($\eta=0.06$ for a standard thin disk around a non-spinning black
hole; a typical assumed value for an accreting black hole is 0.1), we
find $$\varepsilon_{\rm rad} (1+z)= \eta \rho_{\bullet} c^2.$$
$\varepsilon_{\rm rad}$ is the observed energy density in that
radiation now, $\rho_{\bullet}$ is the mean mass density added to the
black holes, and $z$ is the mean redshift of the population. Note that
the result is independent of the assumed cosmology and requires only
that the redshift distribution of the sources be known.

\begin{figure}
\centerline{\includegraphics[width=1.2\columnwidth]{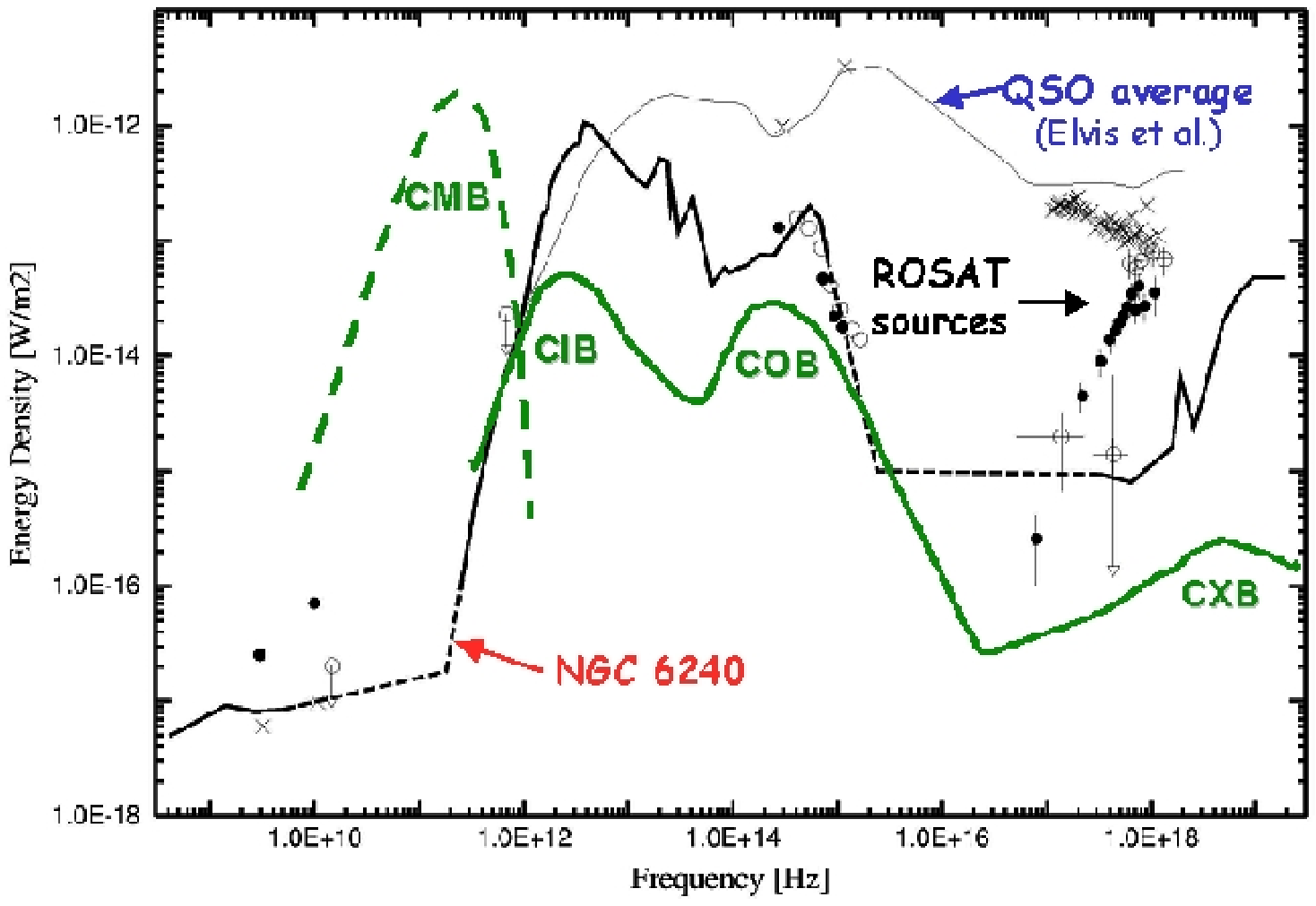}}
\caption{Overview of the intensity of the background radiations
(CIB$=$IRB, COB, and CXB$=$XRB) together with schematic spectral
energy distributions of a nearby obscured AGN, NGC 6240, and of the average
unobscured quasar. (Kindly provided by G. Hasinger; see Hasinger 2002.)}
\end{figure}

$\varepsilon_{\rm rad}$ is determined from either the images and
spectra of the sources themselves or from the background radiation
they produce (Fig. 1.3). A bolometric correction $\kappa$ from the observed band
to the total luminosity is required. $\rho_{\bullet}$ is either
determined from the above equation, giving $\rho_{\bullet}^{\rm AGN}$, or is
measured from local galaxies $\rho_{\bullet}^{\rm direct}$ using the black
hole mass to galaxy bulge velocity dispersion ($M_{\bullet}-\sigma$) 
relation (from Ferrarese \& Merritt 2000 or Gebhardt et al. 2000), the less 
well-known velocity dispersion function for galaxies or a proxy for it, and 
the Hubble constant (in order to determine volumes).  The variables are 
$\eta$, $z$, $H_0$, and $\kappa$.

Various attempts have recently been made to compare the values of
$\rho_{\bullet}^{\rm AGN}$ and $\rho_{\bullet}^{\rm direct}$. In units of
$10^5\Msun\Mpc^{-3}$ for $\rho_{\bullet}$, and adopting $H_0=75\kmpspMpc,$,
Ferrarese (2002) obtained $\rho_{\bullet}^{\rm direct} = 4-5$, whereas Yu \& 
Tremaine (2002) find $\rho_{\bullet}^{\rm direct} = 3\pm0.5$. Using bright 
optical quasars only and $\eta=0.1,$ Yu \& Tremaine (2002) obtain 
$\rho_{\bullet}^{\rm AGN}$ = 2.2. They consider that this agrees well enough 
with their value of $\rho_{\bullet}^{\rm direct}$ that there is no room for 
significant growth by obscured accretion. In other words, they find that the 
mean density in local black holes can be wholly due to radiatively efficient 
accretion in an optically bright (and unobscured) quasar phase.

The result of Yu \& Tremaine (2002) contrasts with that obtained from
the XRB by Fabian \& Iwasawa (1999). In order to use the XRB some
correction has to be made for absorption. This was accomplished by
noting that the mean spectra of unobscured quasars is a power law with
an energy index of unity in the 1--20~keV band. Therefore the minimum
correction is that required to push the XRB spectrum, which is a
power law of index 0.4 in the 2--10~keV band, up to an index of one,
matching at the $EI_E $ peak in the XRB (Fig. 1.4).  This process emphasizes 
the importance of obscured accretion since the unobscured objects dominate
below 1~keV, which is a level 4 times below that of the resultant
absorption-corrected minimum spectrum. In the absorption-corrected
sense there is 3 times more energy density in the obscured objects
than in the unobscured ones. This could imply that obscured accretion
dominates the growth of massive black holes, contrary to the
conclusion of Yu \& Tremaine (2002).

\begin{figure}
\includegraphics[width=0.7\columnwidth,angle=-90]{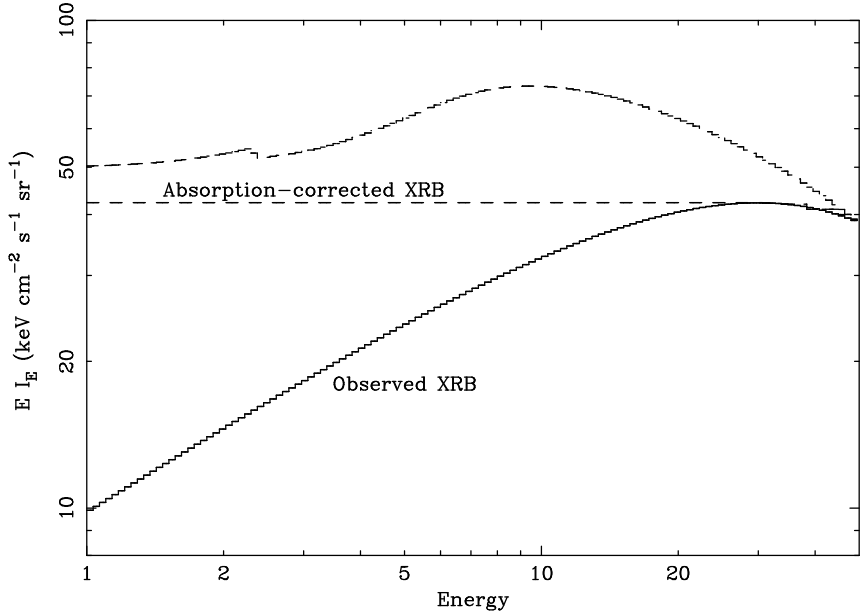}
\caption{Observed $EI_E$ spectrum of the XRB. The spectrum of a typical
unobscured AGN is a horizontal line (energy index of unity). The
minimum correction to the XRB is the horizontal line (dashed)
that matches the XRB spectral peak. Disk reflection (indicated by
top line) could increase this estimate. (From Fabian \& Iwasawa 1999.) }
\end{figure}

The value for $\rho_{\bullet}^{\rm AGN}$ obtained in this way from the XRB
is 6--9 (Fabian \& Iwasawa 1999) and 7.5--16.8 (Elvis et al.  2002). The 
bolometric correction was that relevant for quasars (from the work of Elvis et 
al. 1994), $\kappa_X=30-50,$ $\eta=0.1$, and $z=2$.

The XRB value for $\rho_{\bullet}^{\rm AGN}$ was below the value of
$\rho_{\bullet}^{\rm direct}$ in 1999, when the local quiescent black holes
masses (Magorrian et al. 1998) were about 3--5 times higher than are
found now. If current values of $\rho_{\bullet}^{\rm direct}=3-5$ are used
then it might seem that there is a problem. Elvis et al.  (2002) have
argued that it implies $\eta>0.15$ and therefore that all massive
black holes are spinning rapidly. This could cause some problems with
merger-based galaxy and black hole growth schemes (Hughes \& Blandford 2003).

High radiative efficiency also means that the mass-doubling time
(assuming that the sources are Eddington limited) exceeds 60~Myr,
which could cause problems in growing massive objects from much
smaller seed black holes. Perhaps, however, the plunge region within the
innermost stable orbit is being tapped by the action of magnetic
fields in the disk, thus yielding a higher efficiency without spin
(Gammie 1999; Krolik 1999; Agol \& Krolik 2000). Another possibility
is that many black holes are ejected in mergers, although the required
large fraction seems doubtful.

\section{The Redshift Distribution of Obscured Sources}

The above discussion using the XRB assumed that the evolution of the obscured 
AGNs is the same as that of unobscured quasars, which peaks at $z\approx 2$. 
Recent results from source identifications (Alexander et al. 2001;
Barger et al. 2002; Hasinger 2002; Mainieri et al. 2002; Rosati et al. 2002)
have, however, shown that the obscured objects peak at a
lower redshift of about $z\approx 0.7$. The identification of a complete
sample has not yet been carried out, but, as noted by the above
authors, the results from partial samples already show that there are
many more sources found below redshift 1 than would be expected from
any model based on quasar evolution. This is a very important result.

The immediate effect on the problem in the last section is the drop in
$z$ and also of $\kappa$. Most, but not all, of the X-ray sources now
have absorption-corrected, 2--10~keV luminosities below
$10^{44}\ergps$ and are strictly not quasars. Although this luminosity
distinction is somewhat arbitrary, the key point is that Seyferts have
a 2--10~keV bolometric correction factor $\kappa$ of 10--20,
rather than 30--50 typical of quasars.

It is difficult to determine $\kappa$ for Seyfert galaxies since the
dominant thermal disk emission lies in the rest-frame EUV. Moreover, 
the low redshift of most of the well-studied ones means that the disk
emission, unlike that in quasars, is not shifted into easily observable bands.
I have used results from {\it ASTRO-1, EUVE}, and {\it FUSE}\ and find that
$\kappa=12-18$, using data on Mrk 335 (Zheng et al. 1995), NGC 3783, 
(Krolik \& Kriss 2001) and NGC 5548 (Magdziarz et al. 1998).  

Together, the joint effects of the lower redshift peak and lower
bolometric correction reduce the value of $\rho_{\bullet}^{\rm AGN}$ for
obscured sources by a factor of up to 3. The net result is that
optically bright (unobscured) Type 1 quasars give $\rho_{\bullet}^{\rm AGN} 
\approx 2$, and obscured, Type 2 AGNs now give a similar value,
totaling about 4, again in units of $10^5 \Msun\Mpc^{-3}$. This then
agrees well with the values from local studies of $\rho_{\bullet}^{\rm direct}.$

[Note that the change in $\kappa$ with luminosity means that a lower limit for 
the intrinsic 2--10~keV luminosity of quasars is difficult to
determine at the present time. The origin of the optical definition is
somewhat arbitrary. The properties of objects appear to change around
$L(2-10\keV)=3\times 10^{44}\ergps$, which is the limit used here.]

\subsection{Some Implications for Obscured Sources}

The discovery that obscured AGNs follow a much steeper evolution than 
optically bright quasars means that they are a different population. The lack 
of any unobscured counterpart implies that the obscuration covers a large
part of the sky as seen by the source itself. There can be no simple
torus or geometrical unification picture for these objects. The large
covering fraction of absorbing material may explain why there is
little in the way of an optical/UV narrow-line region seen for some of
the identified X-ray sources (particularly if the absorbing gas is
dusty; see, e.g., Gandhi et al. 2002).

Interestingly, there is another population of objects that does evolve
in a similar way to the obscured X-ray population, namely dust-enshrouded 
starburst galaxies. Chary \& Elbaz (2001) show that distant luminous and 
ultraluminous infrared galaxies seen with {\it ISO}\ evolve very rapidly to 
$z\approx 0.8$. There is some overlap between {\it ISO}\ and 
{\it Chandra}/{\it XMM-Newton}\ X-ray sources in deep images (Wilman, Fabian, 
\& Gandhi 2000; Alexander et al. 2002a; Fadda et al. 2002), and it is
plausible that a subset ($\sim20$\%) of the dusty starburst
galaxies are X-ray detectable Type 2 AGNs. The inner parts of the
starburst itself may, through winds and supernovae, be responsible for
inflating the absorbing gas so that it has a large covering fraction
as seen from the center (Fabian et al. 1998; Wada \& Norman 2002).

It is not yet clear from identification work quite what fraction of
the X-ray sources dominating the XRB are dusty starbursts. Very (and
extremely) red objects are reasonably common counterparts of the
{\it Chandra}\ serendipitous X-ray sources (Alexander et al. 2002b); yet, 
they are a heterogeneous class (e.g., Smail et al. 2002) that includes both 
dusty starbursts and old early-type galaxies.

A further point is that Seyfert galaxies typically operate at about 10\% 
of the Eddington limit. If the XRB objects are similar, then
from the inferred luminosities, the black hole masses are in the
range of $10^6 - 3\times 10^8\Msun$, below that generally
implied for quasars ($\sim 10^8$ to $>10^9\Msun$). Thus, unobscured
accretion seen in optically bright quasars may make most of the black
holes above $3\times 10^8\Msun$ and obscured accretion those at lower
masses.

What we do not know is whether quasars passed though an obscured phase
early on (see, e.g., Fabian 1999), perhaps when their masses were
$10^8\Msun$ or less. As discussed earlier, there are some Type 2
quasars being found in deep X-ray surveys, but not in large numbers.
Also there is little evidence yet for a population of distant
Compton-thick objects ($N_{\rm H}>1.5\times 10^{24}\psqcm$). Of course,
these are likely to be difficult to identify spectroscopically. Wilman
(2002) has provided arguments against any significant Compton-thick
population. There may still be a Compton-thick population to fill in
the $\nu I_{\nu}$ peak in the XRB, but it could well be at {\it low}\
redshift ($z<0.5$). It requires a instruments like the {\it Swift}\ BAT and
EXIST to uncover this population in detail.

Unless the estimates for $\rho_{\bullet}^{\rm direct}$ are revised upward in
the future, there is little room for $\rho_{\bullet}^{\rm AGN}$ in terms of
Type 2 quasars or Compton-thick objects, with a limit being 
\lax $1 \times 10^5 \Msun\Mpc^{-3}$.

\section{Models for the Evolution of the XRB}

The newly discovered redshift distribution (Fig. 1.5) for the obscured AGNs has
prompted some new synthesis models for the XRB in which the obscured
objects have a steeper evolution than the unobscured ones
(Franceschini, Braito, \& Fadda 2002; Gandhi \& Fabian 2003). 

\begin{figure}
\hspace{1.5cm}
\includegraphics[width=.5\columnwidth,angle=90]{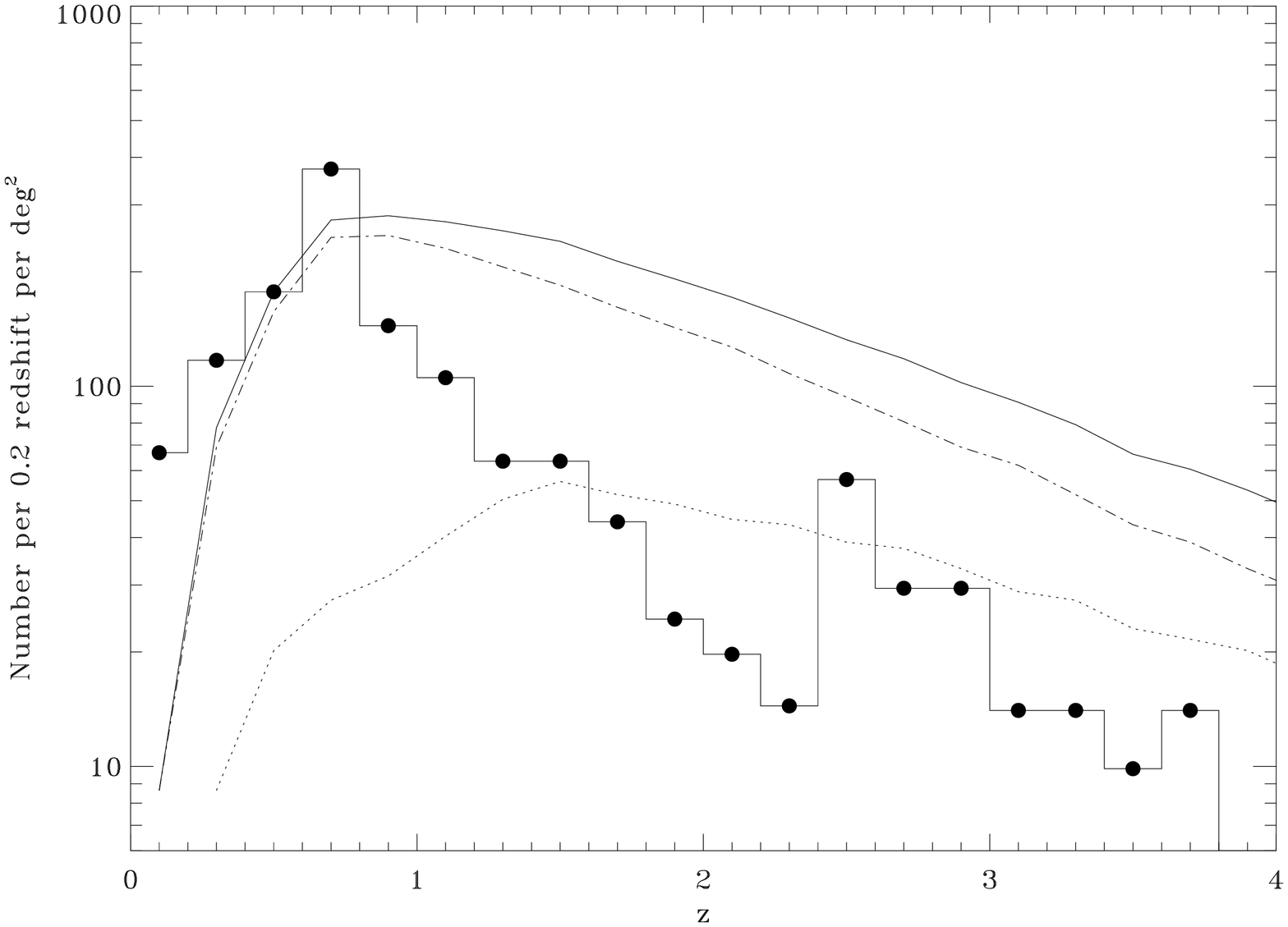}

\hspace{1.5cm}
\includegraphics[width=.5\columnwidth,angle=90]{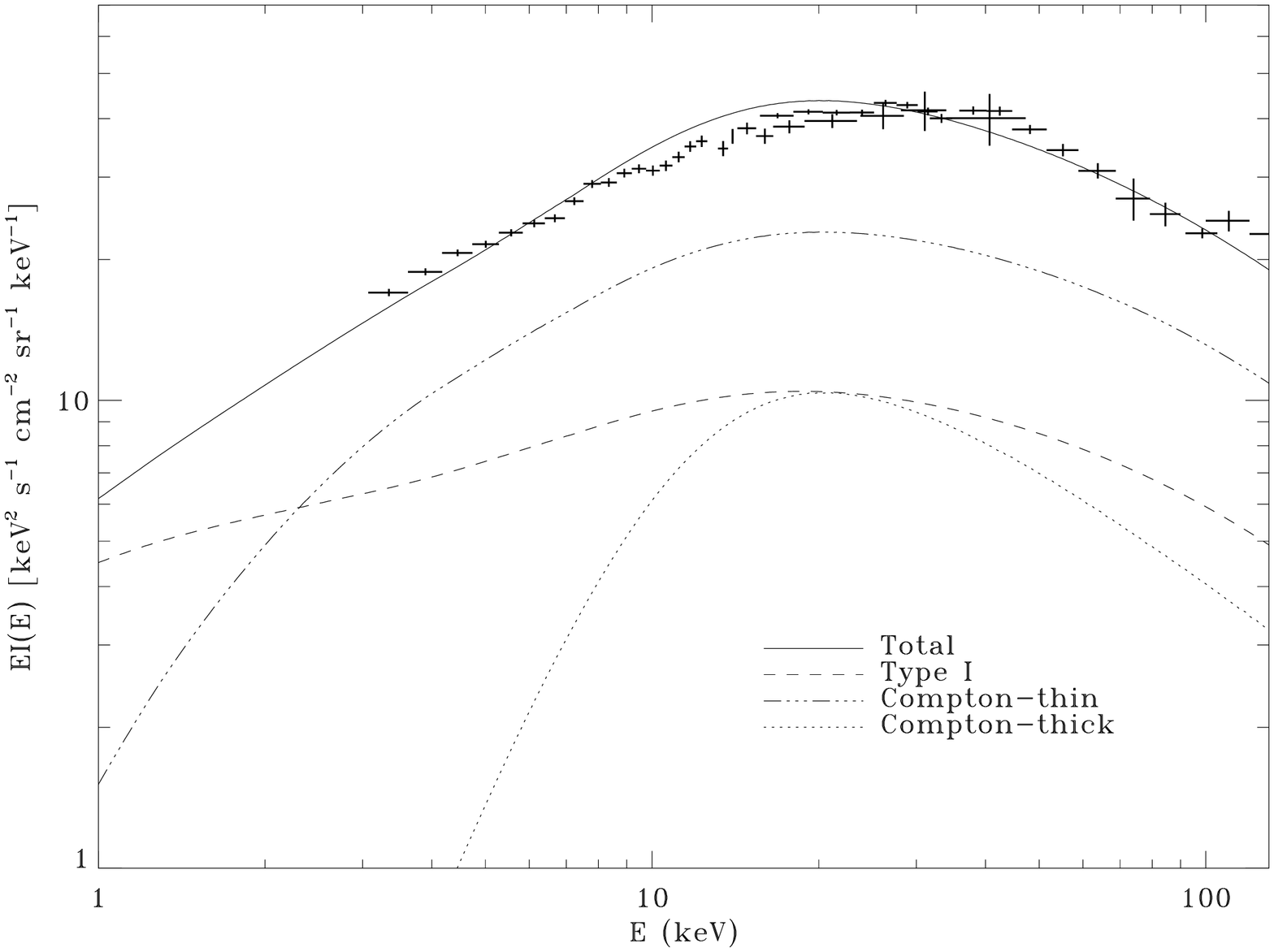}

\hspace{1.5cm}
\includegraphics[width=.5\columnwidth,angle=90]{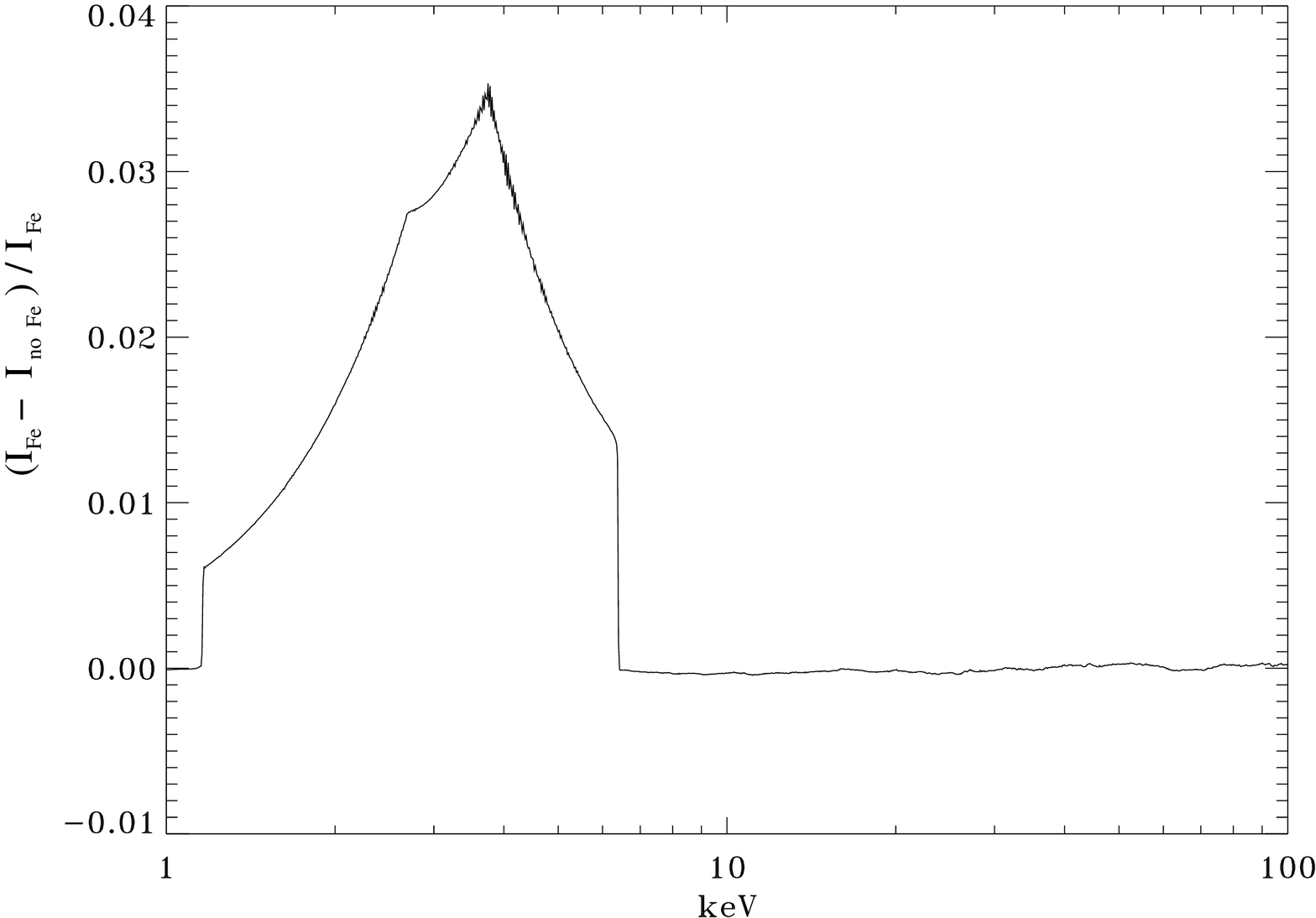}
\caption{{\it Top:}\ The observed serendipitous X-ray source distribution with
redshift (Hasinger 2002), with the predictions from the model of Gandhi
\& Fabian (2003). Type 1 and 2 sources are the dotted and dashed lines, 
respectively. Source identifications are not complete, so more sources may 
fill in above $z=1$. The numbers found below $z=1$ already far exceed the 
predictions of the model by Gilli, Salvati, \& Hasinger (2001).  {\it Middle:}\
Matching the model to the XRB spectrum. It is not clear that a significant 
Compton-thick population is required. {\it Bottom:}\ The spectral residuals 
expected due to the presence of iron fluorescence emission in the sources.}
\end{figure}

A difficulty in making such models is the lack of any X-ray luminosity
functions for Type 2 AGNs. Gandhi \& Fabian (2003) used the $15\, \mu m$ 
{\it ISO/IRAS}\ infrared luminosity function of Xu et al. (1998).
Spectral energy distributions suggest that $L(2-10\keV)\approx L(15\, \mu
m),$ and using that we normalize to the local 2--10~keV X-ray
luminosity function of Piccinotti et al. (1982). A power-law
distribution in column density is assumed. A reasonable fit to
the spectrum of the XRB is obtained if the emissivity due to Type 2
AGNs evolves as $(1+z)^4$ out to $z=0.7$ and then remains flat to
higher redshifts (at least to $z=1.5$). $\rho_{\bullet}^{\rm AGN}$ from this
model is in accord with the values in \S~1.4. The model can also
reproduce well the X-ray source counts, provided that some density
evolution is included.

Gandhi \& Fabian (2003) tested whether the iron emission line produced by 
fluorescence in the absorption process is detectable in the spectrum (Matt \&
Fabian 1994) and found that it should give a small peak in the XRB
spectrum over the 3--4~keV band. Interestingly, there is a bump in the
{\it ASCA}\ spectrum of the XRB at that point (Gendreau 1996), although it is 
not statistically significant. {\it XMM-Newton}\ should do better and could
thereby confirm, in an integral manner, the redshift distribution of the
component sources of the XRB.

\section{Some Comments on Fueling and Obscuration}

Rapid fueling of a black hole requires a plentiful gas supply, which,
if distributed, is likely to coincide with a starburst. Star formation
can churn the gas up through winds and supernovae and make the
covering fraction of cold absorbing gas large (Fabian et al. 1998; Wada
\& Norman 2002). The coincidence between
the {\it ISO}\ dusty starburst population and the obscured Type 2 AGNs should
not then be too surprising. What has yet to be explained is why there
is such a dramatic and rapid decrease in this activity since a
redshift of 0.7 [proportional to $(1+z)^4$], over the last 5~Gyr.

The question is then raised of why quasars are unobscured. It could be
the high luminosity that drives away all nearby gas. Perhaps, as
indicated above, the optically bright quasars are of higher mass
(above $10^8\Msun$) than the typical Type 2 AGNs, or they are closer
to the Eddington limit. Powerful radio galaxies, which have very
luminous and presumably massive nuclei do, however, provide counter
examples; although the radio outbursts may all be young ($<10^7\yr$)
and due to even higher mass black holes, they could be well
sub-Eddington. A significant population of Type 2 quasars may yet
emerge from the complete identification of large, deep X-ray samples.

The Eddington limit for a central black hole will always be
significantly less than the Eddington limit for the galaxy bulge. That
assumes, however, only radiation pressure through electron scattering. A
nuclear wind or radiation pressure acting on dusty cold gas (which has
a much larger absorption cross section than electron scattering) can
make the relevant limiting luminosity for the galaxy bulge more than
the Eddington limit for the nucleus itself (Fabian 1999; Fabian,
Wilman, \& Crawford 2002). Consequently an obscured nucleus can
increase in luminosity and then blow away the obscuring gas and its
own fuel supply (see also Silk \& Rees 1998). This can help relate the
final black hole mass to the mass and potential well of its host
bulge. The tight $M_{\bullet}-\sigma$ relation found locally (Ferrarese
\& Merritt 2000; Gebhardt et al. 2000) does, however, suggest that a
single mechanism is acting throughout the entire mass range.

\section{Contributions to the Far-infrared and Sub-mm Backgrounds}

The X-ray and UV luminosity absorbed in dusty Type 2 AGNs is
re-emitted in the far-infrared. Much work remains to be done, but it is
plausibly reradiated at about $100\, \mu m$ in the rest frame. From an
{\it ISO}\ study Fadda et al. (2002; see also Alexander et al. 2002a) find that
AGNs contribute about 17\% of the infrared background at
$15\, \mu m$. In a $\nu F_{\nu}$ sense this contribution will rise by about
a factor of 2 out to a few $100\, \mu m$ (see, e.g., Crawford et al. 2002),
whereas the IRB rises by about a factor of 10, such that the total
contribution of AGNs to the whole IRB will be 3\%--4\%. 

The AGN fractional contribution to the IRB is highest at the shorter
wavelengths and drops at the longer wavelengths. Only if there is some
as yet unidentified population of Compton-thick objects can this
fraction be much larger and important at long wavelengths, in the
sub-mm. The margin for such a population, given the agreement between
the predicted and observed local black hole densities, is small. The
{\it Chandra}\ medium-deep detection rate for serendipitous SCUBA sources is
low (Bautz et al. 2000; Fabian et al. 2000; Hornschemeier et al. 2000; 
Barger et al. 2001; Almaini et al. 2003). In the 2~Ms CDF-N, Alexander et
al (2003) detect 7 out of 10 bright SCUBA sources and classify 5 as
AGNs. They find luminosities that are Seyfert-like and conclude that
the sub-mm emission is dominated by starbursts.

Fabian \& Iwasawa (1999) determined the IRB contribution to be
2~nW~m$^{-2}~\sr^{-1}$, while Elvis et al. (2002) found
3.6--8~nW~m$^{-2}~\sr^{-1}$; these values are to be compared with a total
integrated intensity of 40~nW~m$^{-2}~\sr^{-1}$ for the IRB. Revising
the bolometric correction factor $\kappa$ down to 15 then makes these
predictions range between $\sim 1-3$~nW~m$^{-2}~\sr^{-1}$. In other
words, AGNs, principally obscured ones, contribute a few percent to the IRB.

Our understanding of the source composition of the IRB will receive an
enormous boost from the imminent launch and operation of {\it SIRTF}.

\section{Summary}

The XRB and X-ray source populations show that there is much obscured
accretion in the Universe. This creates a black hole growth crisis if
the Yu \& Tremaine (2002) analysis stands, unless AGNs are all
particularly radiatively efficient ($\eta>0.15$), perhaps with rapidly
spinning black holes (Elvis et al. 2002). However, if the mean local density of
black holes is between 4 and $5\times 10^5\Msun\Mpc^{-3}$, as
deduced by Ferrarese (2002), then there is consistency for $\eta\approx
0.1$ between the density predicted for both obscured and unobscured
AGNs using the XRB. They contribute roughly equal amounts to the local
mass density of black holes, with obscured accretion contributing most
for black holes with masses below about $3\times 10^8\Msun$ and
unobscured quasars contributing most above that value.

The overall picture is that the most massive black holes (above the
break in the present mass function at about $3\times 10^8\Msun$) are
built earlier (by $z\approx1.5$) than the lower mass ones below the
break (which are built by $z\approx0.7$). This means that the lower
mass ones take about twice as long to assemble. As the Universe ages, the
black holes that remain active are becoming increasingly obscured.

The fraction of the mass density from obscured accretion is lower than
estimated earlier because the sources evolve differently to quasars,
peaking at $z=0.7$ rather than 2. Also their luminosities are in the
range of Seyferts which locally have a lower bolometric correction.
This makes the contribution of obscured accretion to the mass density
in black holes at about 50\%. Uncertainties due to the level
of spin and efficiency for both obscured and unobscured AGNs remain,
and to the exact evolution of complete samples of the X-ray sources.
Also there could be populations of Type 2 quasars and distant
Compton-thick sources yet to be discovered. Such populations can only
make a significant contribution if either optically bright quasars are
found to be rapidly spinning or the local mass density in black holes
has been seriously underestimated.

The covering fraction of the obscured Type 2 AGNs which dominate the
XRB must be very high, or the unobscured fraction would already have
been noticed. The obscuring material probably forms part of a compact,
inner dusty starburst in the host galaxy. The absorbed luminosity is
reradiated in the mid-infrared to far-infrared bands and contributes a few per
cent of the IRB.

Complete Type 2 AGNs samples at all redshifts are urgently needed.
Obscured accretion, which is best studied at X-ray wavelengths, must
be studied in order for the growth and evolution of massive black
holes to be understood.

\vspace{0.3cm}
{\bf Acknowledgements}.
I thank Luis Ho for organizing such an interesting meeting, and Niel
Brandt, Poshak Gandhi, G\"unther Hasinger, Kazushi Iwasawa, and Jeremy
Sanders for help.

\end{document}